\documentclass[twocolumn,showpacs,preprintnumbers,amsmath,amssymb]{revtex4}
\usepackage{graphicx}
\usepackage{dcolumn} 
\usepackage{bm}      

\begin{document}

\preprint{FERMILAB-PUB-11-598-APC}
\title{Transverse modes and instabilities of a bunched beam\\ 
with space charge and resistive wall impedance}
\author{V.~Balbekov}
\affiliation {Fermi National Accelerator Laboratory\\
P.O. Box 500, Batavia, Illinois 60510}
\email{balbekov@fnal.gov} 
\date{\today}

\begin{abstract}

Transverse instability of a bunched beam is investigated with synchrotron 
oscillations, space charge, and resistive wall wakefield taken into account.
Boxcar model is used for all-round analysis, 
and Gaussian distribution is invoked for details.
The beam spectrum, instability growth rate and effects of chromaticity
are studied in a wide range of parameters,
both with head-tail and collective bunch interactions included.
Effects of the internal bunch oscillations on the 
of collective instabilities is investigated thoroughly. 
Landau damping caused by the space charge tune spread is discussed, 
and the instability thresholds of different modes of Gaussian bunch are estimated.

\end{abstract}


\maketitle

%

\section{Introduction}

%

Transverse instability of a bunch in a ring accelerator has been considered 
independently by Pellegrini \cite{PEL} and Sands \cite{SAN} with synchrotron 
oscillations and some internal degrees of freedom of the bunch taken into 
account (``head-tail instability'').
Later~Sacherer have investigated the effect in depth including dependence of 
the bunch eigenmodes on amplitude of synchrotron oscillations (``radial 
modes'' \cite{SAC}). 

Space charge field of the bunch was not taken into account in these pioneer 
works. 
Its role was studied first in Ref.~\cite{MY1}~where it has been shown that 
the space charge tune spread results in Landau damping which suppresses 
many of the head-tail modes, much like other sources of the incoherent tune 
spread.  

However, the last result has been obtained with an assumption that the space 
charge tune shift is significantly less than the synchrotron tune.
A closer examination of the problem in Ref.~\cite{MY2} has led to the 
conclusion that almost all head-tail modes are prone to Landau damping till 
then the space charge tune shift is about twice less then the synchrotron 
tune.
The damping vanishes when the shift becomes more, lower eigenmodes being 
free from the decay first.    
The lowest (rigid) mode is the only universal exception from the rule being 
potentially unstable with any space charge. 
Sometimes 1--2 next modes can demonstrate similar behavior, in dependence 
on the bunch shape.

However, no wakefield was included in the analysis of Ref.~\cite{MY2} in fact,
so the results might be interpreted only as conditions which permit the bunch 
instability but do not determine its characteristics completely. 
Effects of short wakes were actually studied in Ref~\cite{MY3} where the 
instability growth rate has been found at different bunch parameters like 
its length, chromaticity, etc.
Transverse modes coupling instability was considered in the work as well, 
and it has been shown that this more rigorous effect can be caused only by 
positive wake (for comparison, resistive wall creates a negative wake).

A single bunch was investigated in the mentioned articles because only 
short-range forces have been accepted in all the cases.
More general theory is developed in this paper where a bunched beam with 
arbitrary number of the bunches is examined taking into account the space 
charge, intra-bunch and bunch-to-bunch interactions.  
Both kinds of the interactions affect the beam eigenmodes and the instability 
growth rate, although one of them can dominate is specific situation. 
Just from this standpoint the common used terms like ``collective modes 
instability'' or ``head-tail instability'' are treated in the paper. 

The presentation is focused first on the resistive wall instability 
\cite{SES}--\cite{DIC} but the results can be rather easy adapted for other 
known wakefields.
The boxcar model is intensively used to get a general outlook of the problem 
in a wide range of parameters.
More realistic Gaussian bunch is closely examined in a limiting case of low 
synchrotron frequency, although the opposite case of high frequency is also 
invoked to estimate thresholds of Landau damping. 

Both betatron and synchrotron bare oscillations are implied to be linear in 
this paper. 
The nonlinear effects unquestionably require a special investigation, especially 
as an additional factors of Landau damping.  

%

\section{Bunched beam general equations}

%
With space charge and wakefield taken into account, equation of coherent 
betatron oscillations of a bunched beam in the rest frame can be written 
in the form~\cite{MY3}:
%
\begin{eqnarray}
 \left(\frac{\omega}{\Omega_0}-Q_0\right) Y 
+i\,Q_s \frac{\partial Y}{\partial\phi}+\Delta Q\,(Y-\bar Y)         \;\;\,\nonumber\\ 
=\,2\int_\theta^\infty \exp\left[\,i\,(\theta'-\theta)
 \left(\frac{\omega}{\Omega_0}-Q_0-\zeta\right)\right]               \nonumber\\
 \times\;\,q\,(\theta'-\theta)\bar Y(\theta')\rho(\theta')\,d\theta' \hspace{24mm}
\end{eqnarray}
%
where $\theta$~is longitudinal coordinate (azimuth),~$\omega$~is frequency 
of the coherent oscillations, $Q_0$~and $\Omega_0$~are central betatron tune 
and revolution frequency (in the laboratory frame), $\Delta Q(\theta)$~is the 
space charge tune shift averaged over transverse directions, 
$Q_s$~and $\phi$~are synchrotron tune and phase, $\zeta=-\xi/\eta$~is 
normalized chromaticity, and $q(\theta)$~is normalized wakefield which 
specific form will be defined later.  
The bunch shape in the longitudinal phase space $(\theta,u)$~is described by 
a distribution function $F(\theta,u)$~with corresponding linear density:
%
\begin{equation}
 \rho(\theta) = \int_{-\infty}^{\infty} F(\theta,u)\,du 
\end{equation}
%
(normalization of the functions will be specified later).
The function $Y(\theta,u)$~is an average transverse displacement of 
particles located in the point $(\theta,u)$~of the longitudinal space, 
multiplied by the factor $\exp\left[i\,(Q_0+\zeta)\theta\right]$. 
Additional averaging of this function over momentum is denoted as 
$\bar Y(\theta)$~being defined by the expression:
%
\begin{equation}
 \rho(\theta)\bar Y(t,\theta) = 
 \int_{-\infty}^{\infty} F(\theta,u)\,Y(t,\theta,u)\,du 
\end{equation}
%

It is more convenient to represent the right hand part of Eq.~(1) as a sum
over all the bunches preceding the examined one (including all preceding 
turns): 
%
\begin{eqnarray}
 \quad\nu Y_n+i\,Q_s \frac{\partial Y_n}{\partial\phi}+\Delta Q\,(Y_n-\bar Y_n)
 \hspace{29mm}\nonumber\\ 
 \hspace{-5mm}=2\int_\theta^{end} \exp\left[-i\lambda(\theta'-\theta)\right]
 q(\theta'-\theta)\bar Y_n(\theta')\rho_n(\theta')\,d\theta'    \nonumber \\
 \;+\;2\hspace{-2mm}\sum_{m=n+1}^\infty \int\limits_{(m)} 
 \exp\left[-i\lambda\left(\frac{2\pi m}{M}+\theta'-\theta\right)\right]\hspace{15mm}
 \nonumber \\ 
 \,\times\; q\left(\frac{2\pi m}{M}+\theta'-\theta\right)
 \bar Y_m(\theta') \rho_m(\theta')\,d\theta'\hspace{20mm}
\end{eqnarray}
%
where $\;\nu=\omega/\Omega_0-Q_0,\;\lambda=\zeta-\nu \simeq \zeta$,~ 
$M$~is number of bunches in the beam, $\;\rho_m(\theta')=\rho(2\pi m/M+\theta')$,
~and $Y_m(\theta')=Y(2\pi m/M+\theta')$.
Any integral in this expression is taken over one of the bunches which are not 
presumed yet to be identical (in particular, some of them may be empty).
However, periodicity conditions must be satisfied in any case:
%
\begin{eqnarray}
 \rho_{m+M}(\theta)=\rho_m(\theta),\hspace{28mm}\nonumber\\
 \bar Y_{m+M}(\theta)=\bar Y_m(\theta)\exp\left[-2\pi i (Q_0+\zeta)\right]
\end{eqnarray}
%
%

\subsection{Symmetric beam}

%
In subsequent sections we will consider symmetric beams composed of identical 
equidistant bunches: $\rho_n(\theta)=\rho(\theta)$.
It would be checked than that all the solutions of Eq.~(4) fall into
$M$~groups which are known as {\it collective modes}: 
%
\begin{equation}
 \bar Y_n^{(k)}(\theta)=\bar Y(\theta)\exp\left[\frac{2\pi in}{M}(Q_0+\zeta-k)\right]
\end{equation}
%
where $k=1,2,...,M$.
There are {\it the head-tail} modes inside the groups, each  
satisfying the equation:
%
\begin{eqnarray}
 \nu Y+i\,Q_s \frac{\partial Y}{\partial\phi}+\Delta Q\,(Y-\bar Y) \hspace{30mm}\nonumber\\ 
 =\;2\int_\theta^{\theta_0} \exp\left[-i\lambda(\theta'-\theta)\right]
 q(\theta'-\theta)\bar Y(\theta')\rho(\theta')\,d\theta'\nonumber \\
+\;2\;\sum_{m=1}^\infty \exp(-2\pi im\kappa) 
 \int_{-\theta_0}^{\theta_0}\exp\left[-i\lambda(\theta'-\theta)\right]\quad\;\nonumber\\
 \times\;q\,\Big(\frac{2\pi m}{M}+\theta'-\theta\Big)
 \bar Y(\theta)\rho(\theta')\,d\theta'\hspace{22mm}
\end{eqnarray}
%
where $\,2\theta_0\,$~is the bunch length, and 
$\;\kappa=(k-\omega/\Omega_0)/M\simeq (k-Q_0)/M$.
Note that the head-tail modes are not autonomous formations but each of 
them more or less depends on the collective mode which it falls. 

The variable $\tau = \theta/\theta_0$~will be used further as a new 
longitudinal coordinate having a range [--1,1], and the normalization 
condition will be imposed
%
\begin{equation}
 \int_{-1}^1\rho(\tau)\,d\tau = 1.
\end{equation}
%


\section{Resistive wake}

%

Resistive wall instability was predicted first it Ref.~\cite{SES}--\cite{KOL}.
Corresponding wakefield function $q(\theta)$~is negative, and can be presented 
in the convenient form:
%
 \begin{equation}
 q(\theta)=-q_0\sqrt{\frac{2\pi}{\theta}},\qquad
 q_0 = \frac{\alpha r_0N_bR^2}{2\pi\gamma\beta Q_0b_y^3}
 \sqrt{\frac{\Omega_0}{2\pi\sigma}}
 \end{equation}
%
where $N_b$~is number of particles per bunch,  
$r_0=e^2/mc^2$~is classic radius of the particle,
$R$~is the machine radius,
$\sigma$~is specific conductivity of the beam pipe, 
$b_y$~is its semi-height, and $\alpha$~is the pipe form-factor 
($\alpha=1$~for a round pipe).
Then one can rewrite Eq.~(7) in the form: 
%
\begin{eqnarray}
 \nu Y +i\,Q_s \frac{\partial Y}{\partial\phi}+\Delta Q\,(Y-\bar Y) = 
-2q_0\sqrt{\frac{2\pi}{\theta_0}}\exp(i\lambda\theta_0\tau) \nonumber\\
\times
\left[\int_\tau^1 \frac{y(\tau')\,d\tau'}{\sqrt{\tau'-\tau}}
+\hspace{-2mm}\sum_{m=1}^\infty\exp(-2\pi im\kappa)
 \int_{-1}^1\frac{y(\tau')\,d\tau'} {\sqrt{Tm+\tau'-\tau}}\right]\nonumber
\hspace{-5mm}\\
\end{eqnarray}
%
where $T=2\pi/M\theta_0$~is the bunch spacing in terms of 
the variable $\tau$, and the notation is used for a shortness:
$y(\tau)=\bar Y(\tau)\rho(\tau)\exp(-i\lambda\theta_0\tau)$.
Square root in the last integral not much depends  on 
the addition $(\tau'-\tau)$,~~so the expansion into Taylor series 
can be applied resulting in:
%
\begin{eqnarray}
 \nu Y +i\,Q_s \frac{\partial Y}{\partial\phi}  
+\Delta Q\,(Y-\bar Y) \; \simeq\; q_0\exp(i\lambda\theta_0\tau) \hspace{10mm}\nonumber\\
 \times\,\bigg[-2\,\sqrt{\frac{2\pi}{\theta_0}}
 \int_\tau^1 \frac{y(\tau')\,d\tau'}{\sqrt{\tau'-\tau}}+\sqrt{M}
 \Big(V_1(\kappa)\int_{-1}^1 y(\tau')\,d\tau' \nonumber\\
-\,V_2(\kappa)B\int_{-1}^1 y(\tau')(\tau'-\tau)\,d\tau'\Big)\bigg]\hspace{18mm}
\end{eqnarray}
%
where $B=M\theta_0/\pi$~is bunch factor, and
%
\begin{eqnarray}
 V_1(\kappa)=-2\sum_{m=1}^\infty \frac{\exp(-2\pi im\kappa)}{\sqrt m},\\
 V_2(\kappa)=-\frac{1}{2}\sum_{m=1}^\infty\frac{\exp(-2\pi im\kappa)}{m\sqrt m}
\end{eqnarray}
%
Both the functions are periodical, one period of $V_2(\kappa)$~being plotted in Fig.~1. 
Because the function $V_1(\kappa)$~has the singularities, its plot is multiplied 
by aperiodic factor $\sqrt{|\kappa|}$~for a convenience.

It goes without saying that form of these plots depends on the wakefield.
However, in most cases it does not change a general structure of Eq.~(11) 
as well as many subsequent inferences. 

%
\begin{figure}[t!]
\includegraphics[width=90mm]{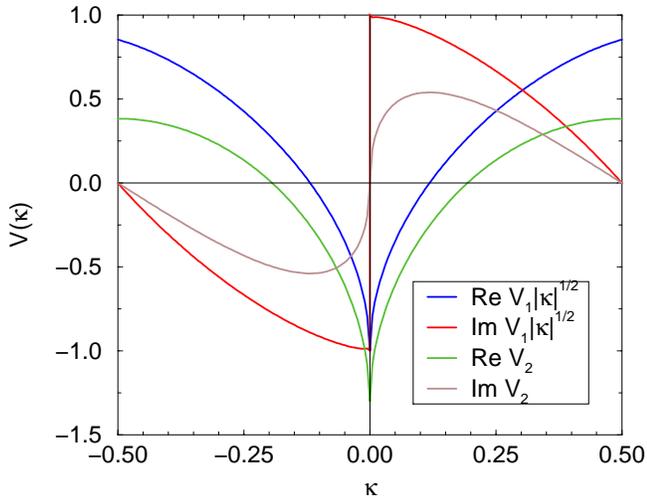}
\caption{Functions $\sqrt{|\kappa|}V_1(\kappa)$~and $V_2(\kappa)$~
(resistive wall wake).}
\end{figure}
%

\section{Boxcar model}

%
The boxcar model is described by the expressions:
%
\begin{equation}
 F = \frac{1}{2\pi\sqrt{1-\tau^2-u^2}},\qquad \rho(\tau) = \frac{1}{2}
 \quad\mbox{at}\quad |\tau|<1
\end{equation}
%
where $u$~is normalized transverse momentum conjugated with the longitudinal 
coordinate $\tau$: $u^2=A^2-\tau^2,\;\;A=\tau_{max}$.
A virtue of this model is that all solutions of Eq.~(11)
are exactly known at $q_0=0$~with any parameters $Q_s$~and $\Delta Q$,~as
it was shown first in Ref.~\cite{SAC}~and developed in detail in 
Ref.~\cite{MY2}. 
This circumstance gives a great possibility to overlook the wakefield 
produced effects.
Required information is shortly reminded below.

All the solutions of Eq.~(11) with $q=0$~are derivable from Legendre 
polynomials $\;\bar Y(\tau)=P_n(\tau),\;n=$~0,\,1,\,2,\,...
At any $n$,~there are $n+1$~different eigenfunctions $Y_{n,m}(\tau,u)$~
with eigentunes $\nu_{n,m}$~where $m=n,n-2,...,2-n,-n$.~
They satisfy the orthogonality conditions:
%
\begin{equation}
 \int\hspace{-2mm}\int FY_{n_1,m_1}Y^*_{n_2,m_2}d\tau du \propto
 \delta_{n_1,n_2}\delta_{m_1,m_2}
\end{equation}
%
\begin{figure}[t!]
\includegraphics[width=90mm]{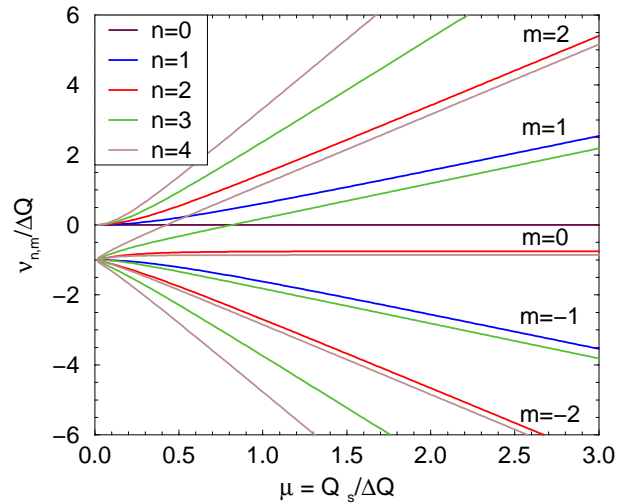}
\caption{Eigentunes of the boxcar bunch}
\end{figure}
%
All the eigentunes are real numbers.
Some of them are plotted in Fig.~2 where the space charge tune shift is used 
for a scaling.
At $\mu\equiv Q_s/\Delta Q=0$,~the eigentunes $\nu_{n,n}$~take start from the
point $\nu=0$~that is $\omega/\Omega_0=Q_0$~(the bare tune). 
Other eigentunes $\nu_{n,m<n}$ start from the point $\nu=-\Delta Q$~
which corresponds the incoherent tune with the space charge included: 
$\;\omega/\Omega_0=Q_0-\Delta Q$.
At $\mu\ll1$,~all eigentunes acquire the additions 
$\delta\nu\propto Q_s^2/\Delta Q$.
In this limiting case, the eigenfunctions have about linear polarization:
$Y_{n,n}$~depends mostly on longitudinal coordinate $\tau$ at any $n$,~
and all other functions $Y_{n,m<n}$~depend mostly on momentum $u$~
(many examples are given in Ref.~\cite{MY2}).
However, these solutions merge together at $\mu\gg1$, forming {\it multipoles} 
$Y_{n,m}(A)\exp(im\phi)$~
with different dependence from synchrotron amplitude.
These {\it radial modes} $Y_{n,m}(A)$~are born by Legendre polynomials of 
powers $n=|m|,\,|m|+2$, etc.
Note that the functions $Y_{n,n}(A)$~are the lowest (minimally oscillating) 
radial modes 
with given $n=m$.
Their eigentunes are $n(n+1)/2\times Q_s^2/\Delta Q$~at small $\mu$~
and $\sim nQ_s$~at large one. 
The functions $Y_{n,m<n}$~are treated usually as higher radial modes.
%

\subsection{Low wake}


The study of the boxcar model is continued in this subsection 
with an assumption that the wake is small enough to apply the perturbation 
methods (applicability of this approximation will be discussed in Sec.~V).
Then $\lambda=\zeta$~in Eq.~(11),~and the additions to the eigentunes 
can be presented in the form:
%
\begin{eqnarray}
 \Delta\nu_{n,m} = q_0\Lambda_{n,m}(\mu)\times\hspace{20mm}\nonumber\\
 \bigg[2\sqrt{\frac{2\pi}{\theta_0}}f_n(\theta_0\zeta)
+\hspace{-1mm}\sqrt{M}\Big(g_n(\theta_0\zeta)V_1(\kappa) 
 \hspace{-1mm}+\hspace{-1mm}ih_n(\theta_0\zeta)V_2(\kappa)B\Big)\bigg]\nonumber
 \hspace{-5mm}\\
 \end{eqnarray}
%
with the coefficients:
%
\begin{eqnarray}
 \Lambda_{n,m}(\mu)=\bigg[\int\int F|Y_{n,m}|^2 d\tau du\bigg]^{-1}   \\
 f_n(\theta_0\zeta) =-\int_{-1}^1y_n^*(\tau)\,d\tau\int_\tau^1 
 \frac{y(\tau')\,d\tau'} {\sqrt{\tau'-\tau}}                          \\
 g_n(\theta_0\zeta)= \Big|\int_{-1}^1 y_n(\tau)\,d\tau\Big|^2         \\
 h_n(\theta_0\zeta)= 2\,\mbox{Im}\int_{-1}^1 y_n^*(\tau)\tau\,d\tau
 \int_{-1}^1 y_n(\tau')\,d\tau' 
\end{eqnarray}
%
Generally, these relations are applicable to a bunch of any form with low wakefield.
However, the boxcar model is the only known case at present which allows to 
investigate the problem in depth because its eigenfunctions $Y_{m,n}(\tau,u)$~
and $y_n(\tau)=P_n(\tau)\rho(\tau)\exp(-i\theta_0\zeta\tau)$
are really known with any $\mu$.~
The results are presented graphically in Figs.~3--7 and commented below.

%
\begin{figure}[t!]
\includegraphics[width=90mm]{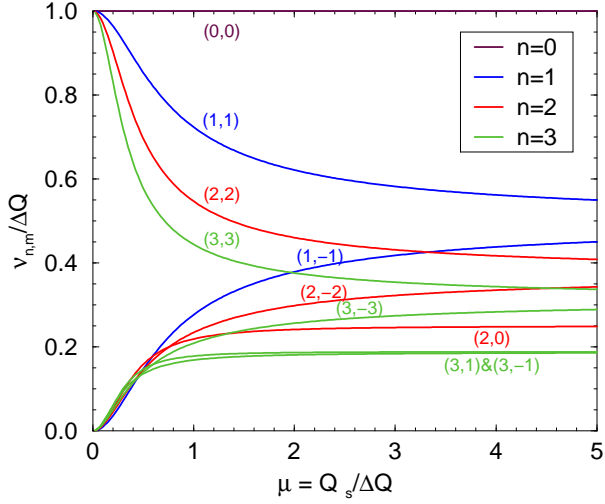}
\caption{Functions $\Lambda_{n,m}(\mu)$~of the boxcar bunch.
Numbers (n,m) are given near each line.}
\end{figure}
%

The coefficients $\Lambda_{n,m}$~depend only on the ratio $\mu=Q_s/\Delta Q$,~ 
and does not depend on the bunch length or chromaticity~(Fig.~3).
They describe a general effect of synchrotron oscillations and space charge 
on transverse coherent motion of the bunches, including possible instability 
growth rate.  
It is seen that the space charge tune shift enhances influence of 
the wakefield on the lowest radial modes $Y_{n,n}$~
but depresses its influence on the higher modes $Y_{n,m<n}$.
It is very understandable result because higher radial modes are 
polarized mostly in $u$-direction at small $\mu$,~so that the global 
bunch displacement at given azimuth and, correspondingly, excited wakefield should be 
relatively small in this limiting case.

%
\begin{figure}[t!]
\includegraphics[width=90mm]{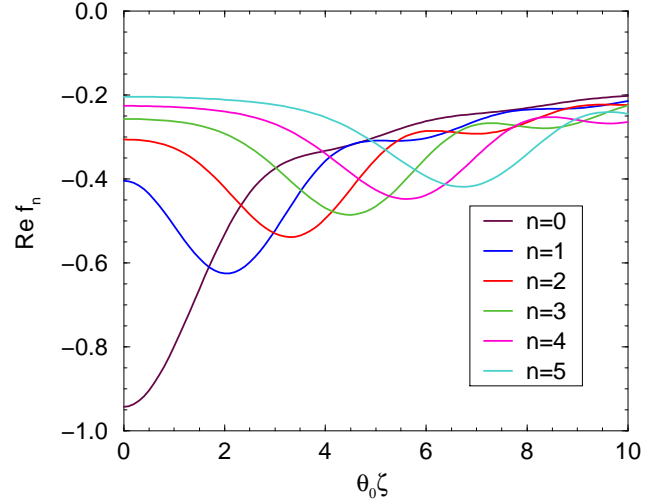}
\caption{Functions Re$\,f_n(\theta_0\zeta)$~of the boxcar bunch. 
The right-hand parts of these even functions are shown. }
\end{figure}
%
\begin{figure}[t!]
\includegraphics[width=90mm]{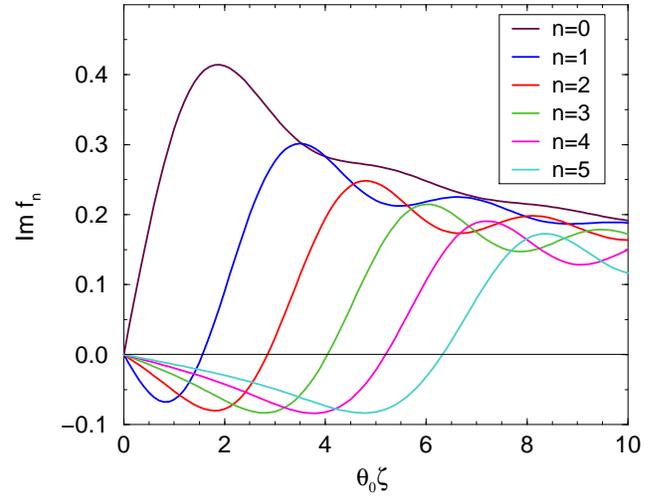}
\caption{Functions Im$\,f_n(\theta_0\zeta)$~of the boxcar bunch.
The right-hand parts of these odd functions are shown.}
\end{figure}
%

In contrast with this, the part of Eq.~(16) in square brackets describes 
effects of the bunch length and chromaticity on different $n$--modes.
There are three terms here which are concerned with different physical effects.

Interaction of particles inside the bunch is described by the coefficients 
$f_n(\theta_0\zeta)$~some of which are plotted in Figs.~4--5.
With non-zero chromaticity, this part is capable to cause the head-tail 
instability of different modes which basic properties were predicted in 
earliest works \cite{PEL}--\cite{SAN}.
It is a single-bunch effect which is proportional 
to the bunch population and does not depend on number of the bunches. 

Second term in Eq.~(16) describes the main effect of the collective 
interaction of the bunches (factor $V_1(\kappa)$~in Fig.~1) including its 
dependence on chromaticity (factor $g_n(\theta_0\zeta))$.
As a rule, the term gives a maximal contribution to the tune shift, 
especially if the beam consists of many bunches. 
Indeed, with $M\gg1$~one can get $\kappa=\sqrt{(k-Q_0)/M}\ll1\;\,$~that is
$\;V_1(\kappa)\simeq(1+\kappa/|\kappa|)/\sqrt{|\kappa|}$\
for the most unstable modes $(k>Q_0)$.~
Then the total contribution of this term to the tune is:
%
\begin{equation}
 \Delta\nu_{n,m}\simeq \frac {q_0 M\Lambda_{n,m}g_n(\theta_0\zeta)}{\sqrt{|k-Q_0|} }
 \left(1+i\,\frac{k-Q_0}{|k-Q_0|}\right)
\end{equation} 
%
With Eq.~(9) for $q_0$~used, it gives the expression: 
%
\begin{equation} 
\frac{\Delta\omega_{nm}}{\Omega_0}
=\frac{\alpha r_0NR\delta(\omega)}{2\pi\gamma Q_0b_y^3} 
 \left(1+i\,\frac{\omega}{|\omega|}\right)\Lambda_{n,m}\,g_n(\theta_0\zeta)
\end{equation}
%
\begin{figure}[t!]
\includegraphics[width=90mm]{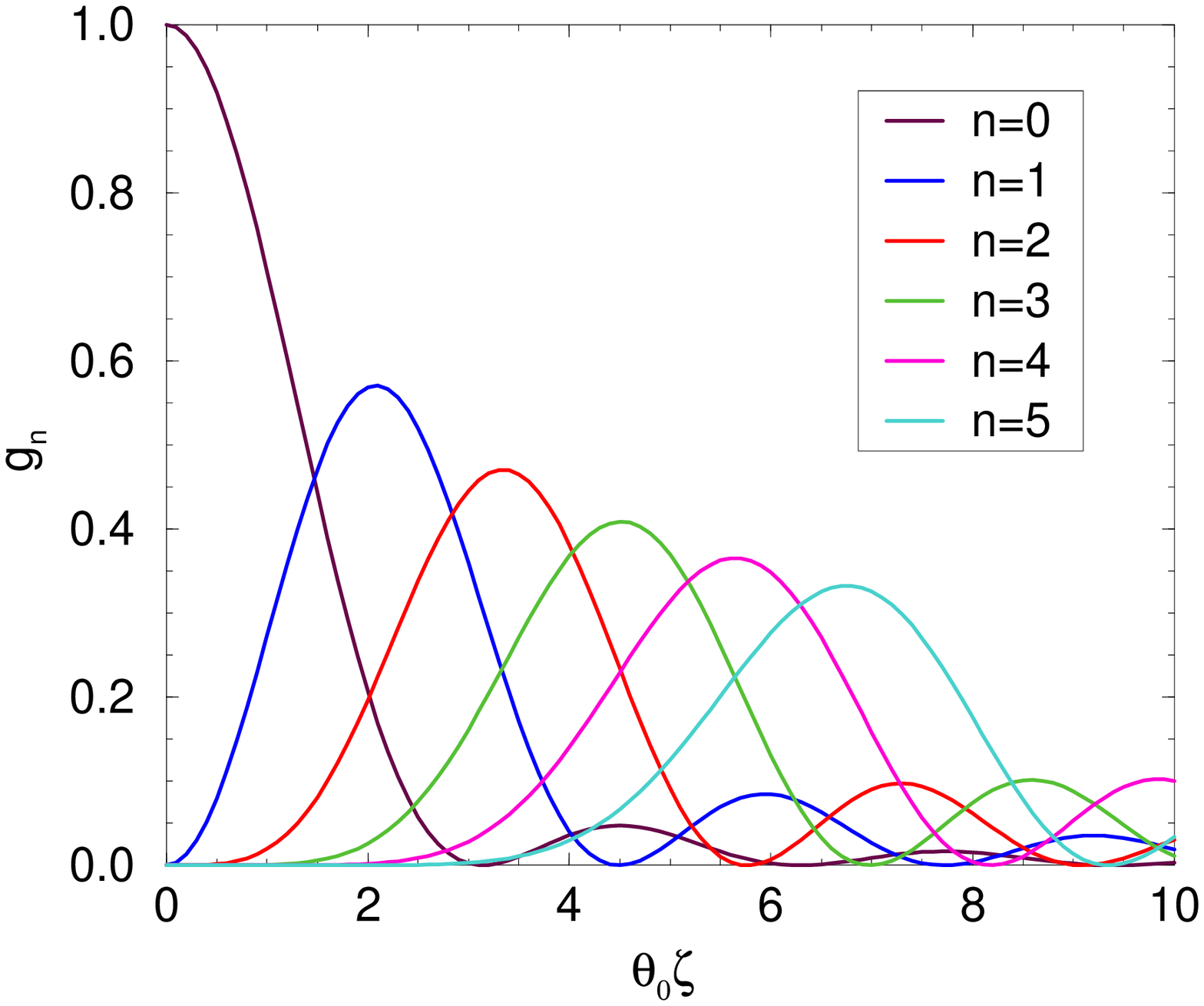}
\caption{Functions $g_n(\theta_0\zeta)$~of the boxcar bunch.
The right-hand parts of these even functions are shown. } 
\end{figure}
%
\begin{figure}[t!]
\includegraphics[width=90mm]{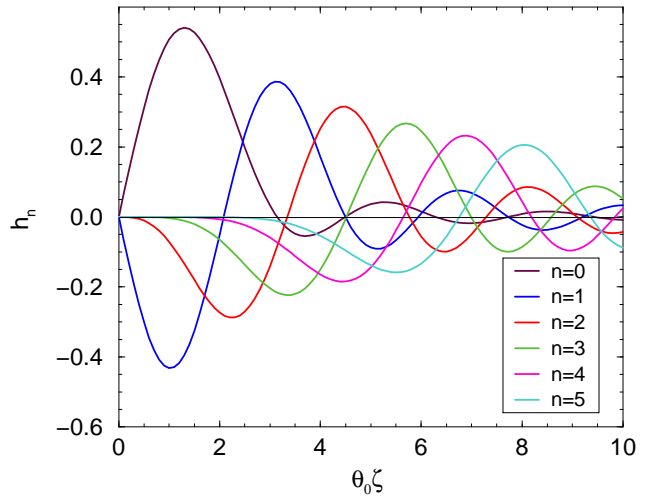}
\caption{Functions $h_n(\theta_0\zeta)$~of the boxcar bunch.
The right-hand parts of these even functions are shown. } 
\end{figure}
%
where $N=MN_b$~is the total beam intensity, $\omega=\Omega_0(k-Q_0)$~
is the coherent frequency in the laboratory frame, and 
$\,\delta(\omega)\,$~is corresponding skin depth.
First part of the formula coincides with well known expression for 
resistive wall instability of a coasting beam \cite{SES}--\cite{KOL}.
The factors $\Lambda_{n,m}$~describe an impact of the bunching upon different 
head-tail modes, including their dependence on synchrotron frequency 
and space charge tune shift. 
Note that the coefficient of the most important rigid mode $\Lambda_{0,0}=1$~
independently on the mentioned parameters, that is the bunching does not 
affect this mode. 

The last term $g_n(\theta_0\zeta)$~in Eq.~(22) describes an influence of 
chromaticity. 
In this regard, it is pertinent to dwell upon the different character of the 
chromatic effects in coasting and bunched beams.

In first case, chromaticity leads to a spread of incoherent betatron 
frequencies which phenomenon can cause Landau damping resulting in total suppressions 
(prevention) of instability.

In contrast with this, average momenta of all the particles are 
equalized in the bunch through synchrotron oscillations.
In such conditions, chromaticity does not provide a systematic tune spread 
and cannot bring about Landau damping at once.
There is no questions that additional slip of betatron phases of particles with respect 
to the coherent field phase affects the interaction and can change the instability growth rate.  
As it follows from Fig.~6, switch of the instability peaks to the higher internal 
modes is the most descriptive result of this. 

However, it is necessary to take into account also other then chromaticity factors 
which can result in  Landau damping. 
In particular, space charge tune spread itself can produce this effect in bunched beams.
As it is shown in Ref.~\cite{MY1}--\cite{MY2}, the higher internal modes 
are more sensitive to this kind of Landau damping.
Therefore, one can expect that the above mentioned shift of the instability
peaks is capable to suppress all internal bunch modes except the rigid 
one which is not prone to this kind of the damping \cite{MY2}. 
However, this problem cannot be solved in frames of the boxcar 
model which ignores this part of the tune spread at all.
Therefore, we postpone its detailed analysis to Sec.~VI where more realistic 
Gaussian distribution will be invoked.  
  
A peculiarity of a single bunch beam is that its instability is possible 
in a restricted region of parameters.  
As it follows from Fig.~1 at $M=1$,~imaginary part of the coefficients 
$V_1$~and $V_2$~is positive at $0<k-Q_0<0.5$.
It means that, without chromaticity, the instability is feasible only if 
betatron tune is located between half-integer and next integer 
(e.g. $Q_0=0.75$~but not 0.25).
This result was first obtained in Ref.~\cite{DIC}~for a short bunch 
where the chromaticity was ignored by the model. 
Actually, it is apparent from Eq.~(16) that the chromaticity is an essential
factor, mainly because it triggers the head-tail interaction.
For example, without chromaticity the bunch is quite stable at $k-Q=-0.25,\;B=0.5$.
However, the rigid mode becomes unstable if $\theta_0\zeta=1$~obtaining the 
growth rate~~Im$\,\Delta\nu_{0,0}\simeq(0.35-0.26)\,q_0=0.09\,q_0$~
(first term in this formula is the head-tail contribution, 
and second one -- turn by turn interaction).
Of course, chromaticity of opposite sigh could prevent such a situation
but higher modes instability would be enforced by this, as it follows 
from Fig.~6.

The last term in Eq.~(16)~is a part of the collective interaction 
which describes the field variation of 
preceding bunches inside the considered one. 
Actually, only the nearest bunch gives a noticeable contribution to this part 
so that the result does not depend on number of bunches, in practice.
Influence of this addition looks much like to the head-tail interaction
which statement can be checked by comparison of Figs.~5 and 7.
However, the effect strongly depends on collective mode as it is 
described by the coefficient $V_2(\kappa)$.
For the example above, its contribution to the rigid mode instability is 
about $-q_0h_0(\theta_0\zeta)/16$~which is less then the ``usual'' head-tail 
effect in order of magnitude. 
Probably, this part of the resistive wake is negligible at any conditions.

%

\section{Low synchrotron frequency}

%

The boxcar model gives a broad outlook of the problem
but maybe it omits some important details being not sufficiently realistic itself.
Therefore another point of view is developed in this section
based on the approach $Q_s\ll\Delta Q$~which is rather characteristic of many proton
machines.
As it is shown in previous section and illustrated by Fig.~3, 
space charge suppresses all the modes $Y_{n,m<n}(\tau,u)$~in this limit.
Therefore the following results are actually concerned only with  
the modes $ Y_{n,n}(\tau)$~which will be denoted later simply as $Y_n(\tau,u)$.
Following Ref.\cite{MY3} with resistive wakefield added, 
one can show that the space part of this function satisfies the equation:
%
\begin{eqnarray}
 U^2(\tau)\bar Y''(\tau)-\Big[\tau+\frac{U^2\Delta Q\rho'(\tau)}
 {(\Delta Q+\nu)\rho}\Big] \bar Y'(\tau)+\frac{\nu(\Delta Q+\nu)}{Q_s^2}
 \bar Y\nonumber\\
 =\frac{q_0\Delta Q \exp(i\lambda\theta_0\tau)}{Q_s^2}
 \;\bigg[-2\sqrt{\frac{2\pi}{\theta_0}}\int_\tau^1 \frac{y(\tau')\,d\tau'}
 {\sqrt{\tau'-\tau}}\hspace{15mm}\nonumber\\
 +\sqrt{M}\left(V_1\hspace{-1mm}\int_{-1}^1 y(\tau')\,d\tau'-V_2B        
 \int_{-1}^1 y(\tau')(\tau'\hspace{-1mm}-\hspace{-1mm}\tau)\,d\tau'\right)\bigg]\;\;\,
\end{eqnarray}
%
where
%
\begin{equation}
  U^2(\tau)=\frac{1}{\rho(\tau)} \int F(\tau,u)u^2du
\end{equation}
%
As it is shown in Ref.~\cite{MY2}~and~\cite{BUR},~ the eigentunes of Eq.~(23)
are $\sim n(n+1)Q_s^2/\Delta Q$~in order of value.
Therefore, with rather small synchrotron frequency and for not very 
high modes, it can be simplified using the approximation 
$|\nu|\ll\Delta Q$~which results in:
%
\begin{equation}
 U^2(\tau)\bar Y''(\tau) = R(\tau)
\end{equation}
%
with 
%
\begin{eqnarray}
 R(\tau) = \left(\tau+\frac{U^2\rho'}{\rho}\right)
 \bar Y'-\frac{\nu\Delta Q}{Q_s^2}\,\bar Y\hspace{30mm}\nonumber\\
 +\;\frac{q_0\Delta Q \exp(i\lambda\theta_0\tau)}{Q_s^2}
 \;\bigg[-2\sqrt{\frac{2\pi}{\theta_0}}\int_\tau^1 \frac{y(\tau')\,
 d\tau'}{\sqrt{\tau'-\tau}}\hspace{15mm}\nonumber\\
+\sqrt{M}\left(V_1\hspace{-1mm}\int_{-1}^1 y(\tau')\,d\tau'-V_2B        
 \int_{-1}^1 y(\tau')(\tau'\hspace{-1mm}-\hspace{-1mm}\tau)\,d\tau'\right)\bigg]\;\;\,
\end{eqnarray}
%
Boundary conditions of this equation are evident from the relation 
$\;U^2(\pm1)=0$~which follows from definition (24) and will be reinforced 
by examples in the subsequent sections.
Therefore, any appropriate solution of Eq.~(25) should satisfy 
the relations:
%
\begin{equation}
  R(\pm1) = 0.
\end{equation}
%
Because Eq.~(25) is linear and uniform, initial conditions 
$\;\bar Y(1)=1,\;R(1)=0$~can be used in practice ~to calculate the function 
$R(\tau)$~everywhere with any trial $\nu$,
and to separate thereafter the eigentunes $\;\nu_n$~
assuring the condition $R(-1)=0$~\cite{MY2}-\cite{MY3}.  
The method is especially effective at $q_0=0$~to determine the basic modes.
In particular, it confirms that Legendre polynomials are solutions 
of the boxcar bunch.  
Generally, it is easy to show that the basic solutions of any bunch 
are regular functions satisfying the orthogonality conditions: 
%
\begin{equation}
  \int_{-1}^1\bar Y_{n_1}(\tau)\bar Y_{n_2}(\tau)\rho(\tau)\,d\tau=\delta_{n_1,n_2} 
\end{equation}
%
Therefore, with enough small $q_0$,~additions to the eigentunes can be 
found by standard perturbation methods which way results in expression like Eq.~(16):
%
\begin{equation}
 \Delta\nu_{n} = q_0\,\bigg[\,2\sqrt{\frac{2\pi}{\theta_0}}f_n +
 \sqrt{M}\Big( g_nV_1 + i\,h_n V_2B\Big)\bigg]
\end{equation}
%
Eqs.~(18)--(20) can be used as well with appropriate eigenfunctions 
$\;y_n(\tau)=\bar Y_n(\tau)\rho(\tau)\exp(-i\theta_0\zeta\tau)$~~
to calculate the factors $f_n,\;g_n,\;h_n$~
All the coefficients $\Lambda_{n,n}=1$~in this case due to normalization 
condition (28).

%

\subsection{Gaussian bunch with low wake}

%
Truncated Gaussian bunch is considered in this subsection 
for a comprehensive investigation of the instability.
Its distribution function is:
%
\begin{equation}
 F=\frac{C}{2\sqrt{2}\sigma}\left(\exp\frac{1-A^2}{2\sigma^2}-1\right)
 \qquad\mbox{at}\qquad A\le1
\end{equation}
%
where the normalizing coefficient $C$~depends on $\sigma$. 
Other involved functions are:
%
\begin{eqnarray}
 \rho(\tau) = C\left(\frac{\sqrt{\pi}}{2}
 \exp(T^2)\,{\rm{erf}}(T)-T\right)  
 \simeq\frac{2CT^3}{3}\left(1+\frac{2T^2}{5}\right)
 \hspace{-5mm}\nonumber\\
\end{eqnarray}
%
\begin{figure}[b!]
\includegraphics[width=90mm]{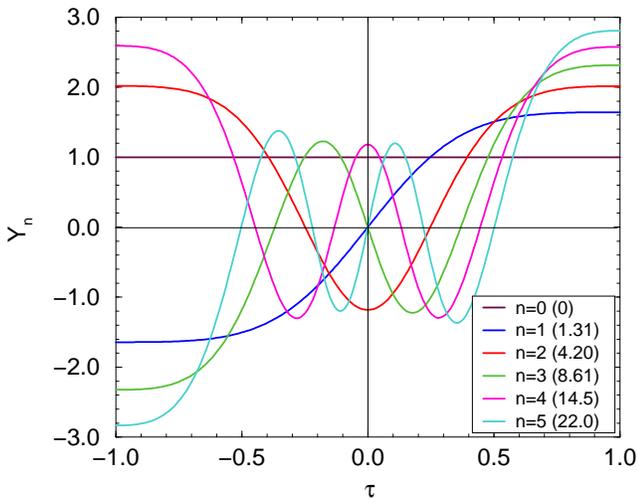}
\caption{Normalized eigenfunctions of truncated Gaussian bunch (0th--5th modes).
Corresponding eigennumbers are given in the parentheses. }
\end{figure}
%
and
%
\begin{equation}
 U^2=\sigma^2\left(1-\frac{2CT^3}{3\rho}\right)
 \simeq \frac {2\sigma^2T^2} {5}  
\end{equation}
%
where $T^2=(1-\tau^2)/(2\sigma^2)$,~and approximate expressions 
at $|\tau|\simeq 1$~are added for references.

The case $\sigma=1/3$~($3\sigma$~truncation, $C=0.016$) is actually 
considered below.
Six basic normalized eigenfunctions of the bunch are shown in Fig.~8.
Their eigentunes have a form $\;\nu_n=\alpha_nQ_s^2/\Delta Q_c$~with
the coefficients $\alpha_n$~which are also presented in Fig.~8 in the brackets
(for comparison: $\alpha_n=n(n+1)/2$ for the boxcar model).
Here and further, the subindex $c$~marks the bunch center.
%
\begin{figure}[t!]
\includegraphics[width=90mm]{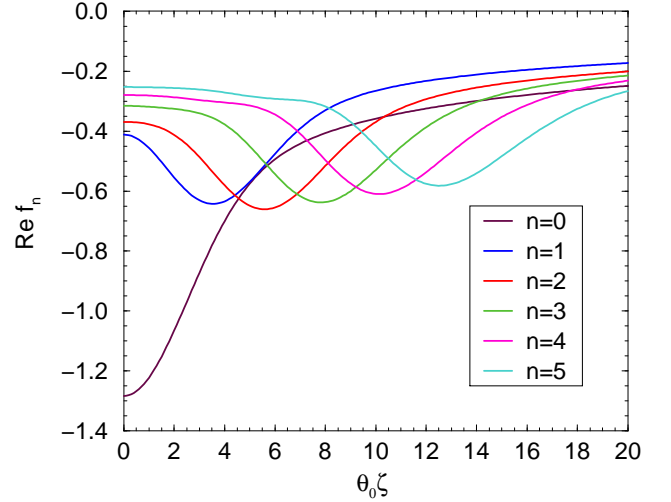}
\caption{Functions Re$\,f_n(\theta_0\zeta)$~of truncated Gaussian bunch. 
The right-hand parts of these even functions are shown. }
\end{figure}
%
\begin{figure}[t!]
\includegraphics[width=90mm]{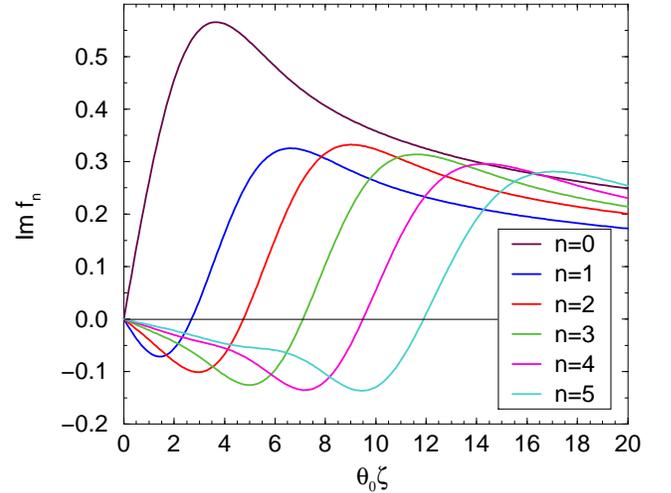}
\caption{Functions Im$\,f_n(\theta_0\zeta)$~of truncated Gaussian bunch.
The right-hand parts of these odd functions are shown.}
\end{figure}
%

The coefficients $f_n,\;g_n$~and $g_n$ are plotted in Figs.~9--12
which look much like Figs.~4--7 of the boxcar model. 
Of course, it is necessary to take into account that the Gaussian bunch 
has less rms length in comparison with the boxcar one 
(1/3 instead $1/\sqrt 3$), so the Gaussian plots should be proportionally wider. 
The absence of secondary oscillations is well explicable because of 
more smooth bunch shapes.
With these reservations, one can assert that the boxcar model 
provides an adequate description of the bunch coherent instability,
at least within the limit of low synchrotron frequency. 
%
\begin{figure}[t!]
\includegraphics[width=90mm]{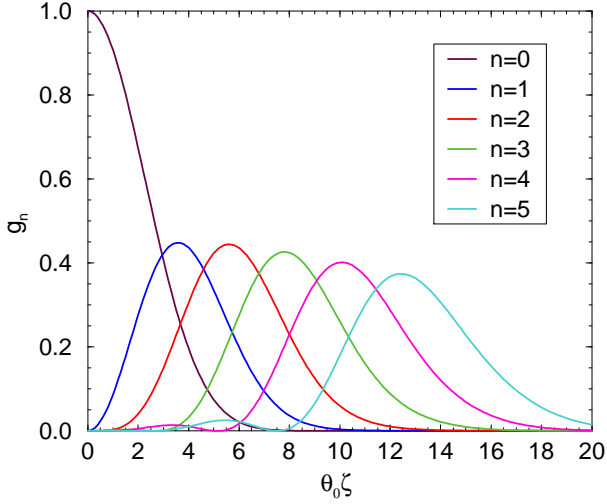}
\caption{Functions $g_n(\theta_0\zeta)$~of truncated Gaussian bunch.
The right-hand parts of these even functions are shown. } 
\end{figure}
%

%

\subsection{Expanded low wake approach}

%

Formally, relation (29) is applicable if the spectrum shifts $\Delta\nu_{n}$~
are small in comparison with distances between the basic spectrum lines which are 
$$
 \nu_{n+1}-\nu_n = (\alpha_{n+1}-\alpha_n)\frac{Q_s^2}{\Delta Q_c}
 \simeq (n+1)\frac{Q_s^2}{\Delta Q_c}
$$
Therefore, condition of applicability of Eq.~(29) is for the lowest mode:
%
\begin{equation}
 |\Delta\nu_0| \ll \frac{Q_s^2}{\Delta Q_c}
\end{equation}
%
The left-hand part of this expression is close to the instability growth rate 
which is essentially less of 1 probably in all practical cases 
(e.g. $|\Delta\nu_0|<0.1$). 
However, the right-hand part can be still less, for example $0.05^2/0.25=0.01$.
The example demonstrates that a violation of condition (33) 
is quite possible occasion, 
especially when collective modes instabilities of a multi-bunch beam are examined.
Therefore, we consider this important case in greater detail, 
without the assumption that the multi-bunch contribution is small.

As a first step, we need to solve Eq.~(25) at $V_1=0$~to know
the basic modes of this case. 
Let us denote corresponding eigenfunctions and eigentunes as $\Upsilon_n$~
and $\upsilon_n$.
Then solution of the total equation can be presented in the form:
%
\begin{eqnarray}
 \bar Y(\tau) = q_0V_1(\kappa)\sqrt{M}
 \int_{-1}^1\bar Y(\tau')\rho(\tau')\exp(-i\zeta\theta_0\tau')\,d\tau'\hspace{-5mm}
\nonumber\\
 \times\;\sum_{n=0}^\infty \frac{e_n\Upsilon_n(\tau)}{\nu-\upsilon_n}\hspace{25mm}
\end{eqnarray}
%
where $e_n$~are coefficients of the expansion:
%
\begin{equation}
 \exp(i\lambda\theta_0\tau)=\sum_{n=0}^\infty e_n\Upsilon_n(\tau)
\end{equation}
It immediately results in the dispersion equation:
%
\begin{equation}
 1 = q_0V_1\sqrt{M}\sum_{n=0}^\infty  \frac{e_n}{\nu-\upsilon_n}
 \int_{-1}^1\Upsilon_n(\tau)\rho(\tau)\exp(-i\zeta\theta_0\tau)\,d\tau 
\end{equation}
%
\begin{figure}[t!]
\includegraphics[width=90mm]{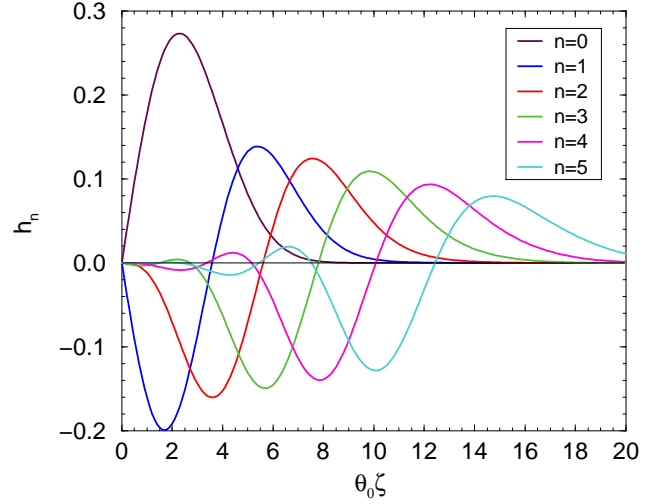}
\caption{Functions $h_n(\theta_0\zeta)$~of truncated Gaussian bunch.
The right-hand parts of these odd functions are shown. } 
\end{figure}
%
In principle, new eigenfunctions $\Upsilon_n(\tau)$~and eigentunes 
$\upsilon_n$~could be found by the same method which was used for $\bar Y_n(\tau)$.
However, it would be a more difficult problem to calculate the coefficients 
$e_n$~because the functions $\Upsilon_n(\tau)$~are not orthogonal, in contrast 
with $\bar Y_n$.
Therefore we turn back to the approximation $\Upsilon=\bar Y$~which is certainly
acceptable at low intra-bunch interaction and does not violate 
the overall structure of Eq.~(36).
It results in the dispersion equation
%
\begin{equation}
 1 = \sum_{n=0}^\infty  \frac{q_0V_1\sqrt{M}g_n(\theta_0\zeta)}
 {\nu-\nu_n-q_0 \left(\sqrt{8\pi/\theta_0}\,f_n +ih_n V_2B\sqrt{M}\right)}
\end{equation}
%
with the same coefficients $f_n,\,g_n,\,h_n$~as before (Figs.~9--12 
for Gaussian bunch).
With an additional condition 
%
\begin{equation}
 q_0|V_1|\sqrt{M} \ll\frac{Q_s^2}{\Delta Q_c}
\end{equation}
%
the low wakefield approximation is totally satisfied, and the equation gives 
the same result as the earlier Eq.~(29).

Another easy case is zero chromaticity when sum (37) holds the only term $n=0$~
because $g_n(0)=\delta_{n,0}$.
It means that solely the rigid internal mode can be excited without chromaticity.
Because it can appear inside any collective mode, 
the resulting tune shift is:
%
\begin{equation}
 \frac{\Delta\omega}{\Omega_0} = q_0\left[2\sqrt{\frac{2\pi}{\theta_0}}f_0(0)
+\sqrt{M}\,V_1\left(\frac{k-Q_0}{M}\right)\right]
\end{equation}
%
This expression formally coincides with Eq.~(29) at $\zeta=0$,~
but can be applied at any value of the coefficient $V_1(\kappa)$.

Generally, series (37) contains restricted number of summands 
which conclusion follows from Figs.~(6) and (11).
In particular, one can see that the terms $n=0$~and 1 give major contributions at 
$|\theta_0\zeta|<\sim3$.  
Eq.~(37) has two actual solutions in this case, at least one of them being unstable. 
In the extreme case when inverse of Eq.~(38) inequality is fulfilled, one of the 
eigentunes is real, and another tune is:
%
\begin{eqnarray}
 \frac{\Delta\omega}{\Omega_0} \simeq q_0V_1(\kappa)\sqrt{M}\;
 \big[g_0(\theta_0\zeta)+g_1(\theta_0\zeta)\big] \hspace{23mm}\nonumber\\
 \simeq \frac{\alpha r_0NR\delta(\omega)}{2\pi\gamma Q_0b_y^3} 
 \left(1+i\,\frac{\omega}{|\omega|}\right)
 \big[g_0(\theta_0\zeta)+g_1(\theta_0\zeta)\big]\quad
\end{eqnarray}
%
Rather weak dependence of this expression on chromaticity engages an attention.
In relative units, the addition is 1 at $\theta_0\zeta=0$~and about 0.8 
at $\theta_0\zeta=3$. 
However, it should be mentioned again that the low synchrotron frequency limit
is considered here. Role of this factor is discussed in the next section. 

%

\section{The instability threshold}

%

It could be concluded from previous analysis that the boxcar is 
a quite adequate model for investigation of the bunched beam instability, 
and only minor and almost obvious changes are needed for 
more realistic distributions. 
However, it would be a premature conclusion because {\it the space charge 
tune spread is ignored at all in the boxcar model}.
Meanwhile, at certain situations the spread can cause Landau damping 
and suppress many unstable modes of a real bunch,
as it has been shown in Ref.~\cite{MY1}--\cite{MY2}.
The rigid intra-bunch mode is the only occasion when this mechanism does not work and 
cannot prevent instability at any combination of parameters.

Unfortunately, at present the problem is adequately covered only in the limiting cases 
$\mu\gg1$~and $\mu\ll1$.
In first case, this kind of Landau damping really
works and suppresses almost all internal modes \cite{MY2}. 
However, the mechanism is turned off in opposite limiting case which 
was just the subject of previous section.
The only conclusion can be done from these facts: this stabilization mechanism 
has a threshold character and actually begin to work at $\mu>\sim1$.
The goal of this section is to get more exact estimation of the thresholds.
Because the space charge will be treated here as a promoting and dominating factor,
the wakefield is excluded from the analysis. 

It should be reminded preliminary that Landau damping arises 
when coherent frequency penetrates rather deeply in a region 
of incoherent tunes of the system. 
Then the particles which own tunes are located below or above the coherent tune
could be exited in contra-phases by the coherent field, transforming its 
energy to the incoherent form (beam heating).
Such a mechanism affects the beam decoherence and creates the instability 
threshold.

Well known practical recommendation follows from this statement 
for coasting beams: incoherent tune spread should exceed the space 
charge tune shift to avoid the instability.
In principle, a wake field (e.g. the resistive wall contribution) affects this 
criterion, however, its influence is small in practice 
if the space charge dominates in the impedance budget.
The last is just the case of our study.

An additional important property of bunched beams is that, 
with a coherent frequency $\omega$, the particles undergo an influence 
of harmonics of frequencies $\omega+j\Omega_s$~where $\Omega_s$~
is synchrotron frequency, and $j$~is integer. 
An intense energy transfer is possible if any of these frequencies falls 
in the incoherent region.
Fast look in Fig.~2 reveals that harmonics $j=m$~have the most chances
to do this.
Therefore, more informative graph can be obtained from Fig.~2 by a
transformation of each curve $\nu_{n,m}(\mu)$~to the $\nu_{n,m}-m\mu$.  
The results are presented in Fig.~13 by the solid lines of the same color
as in Fig.~2.

%
\begin{figure}[b!]
\includegraphics[width=90mm]{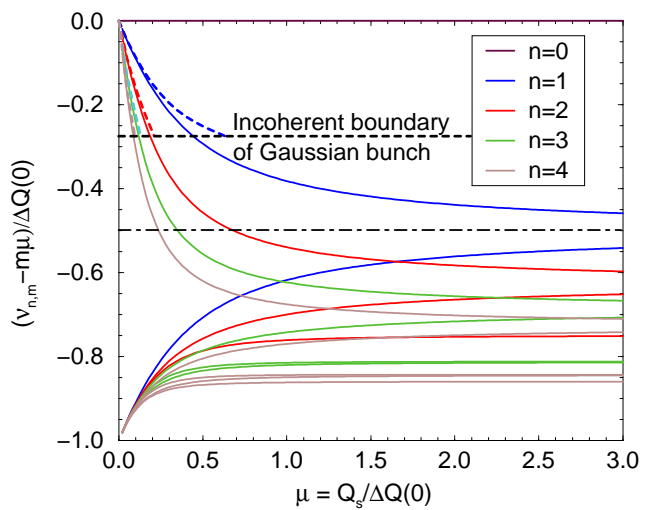}
\caption{Transformed eigentunes: solid lines -- boxcar model;
dashed lines - Gaussian bunch with $3\sigma$~truncation.} 
\end{figure}
%

Averaged over synchrotron phases incoherent tunes of the truncated Gaussian bunch lie 
in the region $-1<\nu/\Delta Q_c<-0.274$. 
Drawing corresponding boundary line in Fig.~13 and assuming that the coherent tunes 
of Gaussian bunch have about the same behavior as in the boxcar model, 
one can make the conclusions:
(i) the lowest (rigid) mode (0,\,0) is unstable with any $\mu$;
(ii) the higher modes $(n,\,n)$~can be unstable at $\mu<\sim0.5$;
(iii) the more is $n$~the less is corresponding threshold of $\mu$;~
(iv) all higher radial modes like $(n,\,m<n)$~are stable in any case.

These conclusions are in a reasonable agreement with results of Ref.~\cite{MY2},
according which only the rigid mode (0,0) of Gaussian bunch is unstable 
at $\mu\gg1$.~
One can anticipate from Fig.~(13) that thresholds of all other modes 
are located at $\mu<0.5$.
It is a region where the low $\mu$~approximation could be fitted to refine 
the thresholds of Gaussian bunch.
To accomplish this, we consider Eq.~(23) with $q_0=0$~but without the additional 
simplification $\nu\ll\Delta Q_c$.
Correspondingly, boundary conditions (27) should be changed by the following one:
%
$$
 Y'(\pm1)=\frac{\nu^2}{Q_s^2}
$$
%
Obtained eigentunes are plotted in Fig.~(13) by dashed lines above 
the Gaussian incoherent boundary. 
The crossing points which are the expected thresholds 
of the head-tail modes are presented in Table 1.  
%
\begin{table}[t!]
\caption{Instability thresholds of Gaussian bunch
with $3\sigma$~truncation (lower radial modes (n,n)).
}
\begin{ruledtabular}
\begin{tabular}{||c|l|l|l|l|l||}
 $n$              &~0&~1&~2&~3&~4\\
\hline
 $ Q_s/\Delta Q_c<$ & $\infty~~~$ &0.63~~~& 0.21~~~& 0.13~~~& 0.095~~\\
\end{tabular}
\end{ruledtabular}
\end{table}
%

Distributions with more abrupt bunch boundaries demonstrate similar behavior
but have higher thresholds.
For example, three modes of parabolic bunch retain the stability 
at $\mu\rightarrow\infty$~\cite{MY2}.  
The boxcar bunch is an extreme case which is unaffected by this sort of damping at all. 

An important conclusion follows from this or similar table,
concerning an influence of chromaticity in real conditions.
As might be expected from Figs.~6 and 11, an increase of chromaticity should 
not affect drastically the instability growth rate because its main effect is 
a simple switch of the bunch oscillations from a lower internal mode to a 
higher one.
For example, it follows from Fig.~11 that the most unstable internal modes of
Gaussian bunch are: $n=1$~at $|\theta_0\zeta|=4$, and 
$n=2$~at $|\theta_0\zeta|=6$,~
in both cases the instability growth rate being about 0.45 in used relative 
units (the head-tail contribution is neglected in the estimations 
because it is reasonably small in a multi-bunch beam).
However, now we must take into account that these results were obtained at $\mu=0$.
Let us consider the case $\mu=0.4$~as an another example.
Then the modes $n\ge2$~cannot appear being suppressed by Landau damping.
Generally, only 0th and 1st internal modes could be unstable in this case, 
and chromaticity $|\theta_0\zeta|\simeq 10$~is sufficient to suppress both of 
them that is to reach a total beam stability.   

Of course, Table 1 is only an estimation of the thresholds 
because approximate Eq.~(23) lies in its foundation.
It would be a good idea to solve more general Eq.~(7) with arbitrary 
$Q_s$~and $\Delta Q_c$,~and to use the results for exact 
instability thresholds of realistic bunches.
Note that all eigennumbers of this equation are real. 
Therefore, the lack of regular solutions with real $\nu$~at some combination 
of synchrotron frequency and space charge tune shift would be a sign of 
Landau damping. 
However, the boxcar model is the only solved case at present. 
%

\section{Summary and Conclusion}

%
Transverse instability of a bunched beam is studied 
with synchrotron oscillations, space charge, and wakefield taken into account.
Resistive wall wakefield is considered as the most universal and practically 
important case. 
However, many results have a common sense and, with small changes, 
can be adapted to other wakes.
Boxcar model of the bunch is extensively used in the paper 
to get a general outline of the problem in wide range of parameters.
A realistic Gaussian distribution is invoked in some cases  
for comparison and more detailed investigations of important problems 
like Landau damping.

Eigenfunctions and eigentunes of the beam are investigated with both 
intra-bunch and inter-bunch interactions taken into account.
Contributions of the interactions to the instability growth rate are studied over
a wide range of the parameters, including effects of the bunch length 
and chromaticity.   
It is shown that known head-tail and collective modes instabilities are 
the extreme cases when one type of the interaction certainly dominates. 
However, an essential influence of the intra-bunch degrees of freedom on 
the collective instabilities is especially marked and investigated in detail.
In particular, it is shown that a variability of the internal bunch modes 
explains why the instability growth rate depends on the bunch parameters 
including space charge tune shift, synchrotron tune, bunch length, 
and chromaticity.   

It is emphasized that the space charge tune spread can cause 
Landau damping and suppress the instability  
(other than  the space charge sources of the tune spread are not 
considered in the paper). 
The phenomenon appears at rather large ratio of synchrotron frequency  
to the space charge tune shift, lower internal modes obtaining the stability 
at larger the ratio.
Several modes of Gaussian bunch are considered in the paper, and their 
thresholds are estimated by comparison of the limiting cases.

%
\section{Acknowledgments}
FNAL is operated by Fermi Research Alliance, LLC under contract 
No.DE-AC02-07CH11395 with the United States Department of Energy. 
%

\end{document}